\documentclass{llncs}
\usepackage{algorithm}
\usepackage{epsfig,psfrag}

\newcommand{\m}{m}

\newcommand{\D}{D}

\newcommand{\carr}{\mathbf{c}}
\newcommand{\cmax}{c^*_{\max}}

\newcommand{\bidgt}{\succ}
\newcommand{\bidlt}{\prec}
\newcommand{\argmax}{\mathrm{arg} \max}

\newcommand{\junk}[1]{}
\newcommand{\eqref}[1]{(\ref{#1})}
\newcommand{\defeq}{:=}

\newenvironment{proofof}[1]{\noindent{\em Proof of #1:}}{\hfill\qed}
\newenvironment{proofsketchof}[1]{\noindent{\em Proof Sketch of #1:}}{\hfill\qed}
\newenvironment{proofsketch}{\noindent{\em Proof sketch:}}{\hfill\qed}
\newenvironment{myproof}{\begin{proof}}{\hfill\qed\end{proof}}

\newenvironment{mechanism}[1]{
\begin{center}
\addtolength{\tabcolsep}{5pt}

\begin{minipage}{4.5in}
\begin{tabular}{|p{4.4in}|}\hline
{\bf #1}\\ \hline
}{
\end{tabular}
\end{minipage}
\addtolength{\tabcolsep}{-5pt}

\end{center}
}

\makeatletter
\renewcommand{\section}{\@startsection{section}{1}{0pt}{-12pt}{5pt}{\large\bf}}
\renewcommand{\subsection}{\vspace*{-.1in}\@startsection{subsection}{2}{0pt}{-12pt}{-5pt}{\normalsize\bf}}
\renewcommand{\subsubsection}{\vspace*{-.1in}\@startsection{subsubsection}{3}{0pt}{-12pt}{-5pt}{\normalsize\bf}}
\makeatother

\title{A Truthful Mechanism for Offline Ad Slot Scheduling}

\author{Jon Feldman\inst{1}
\and S. Muthukrishnan\inst{1}
\and Evdokia Nikolova\inst{2}
\and Martin P\'al\inst{1}
}

\institute{Google, Inc. Email: \email{\{jonfeld,muthu,mpal\}@google.com}
\and Massachusetts Institute of Technology\thanks{This work was done while the author was visiting Google, Inc., New York, NY.}.  Email: \email{nikolova@mit.edu}}

\begin{document}

\maketitle
\begin{abstract}
We consider the {\em Offline Ad Slot Scheduling} problem, where
advertisers must be scheduled to {\em sponsored search} slots during a 
given period of time.  Advertisers specify a budget constraint, as well
as a maximum cost per click, and may not be assigned to more than
one slot for a particular search.
\smallskip

We give a truthful mechanism under the utility model where bidders try
to maximize their clicks, subject to their personal constraints.  In
addition, we show that the revenue-maximizing mechanism is not
truthful, but has a Nash equilibrium whose outcome is identical to our
mechanism.  As far as we can tell, this is the first treatment of
sponsored search that directly incorporates both multiple slots
and budget constraints into an analysis of incentives.

\smallskip

Our mechanism employs a descending-price auction that maintains a
solution to a certain machine scheduling problem whose job lengths
depend on the price, and hence is variable over the auction. 
The price stops when the set of bidders that can
afford that price pack exactly into a block of ad slots, at which
point the mechanism allocates that block and continues on the
remaining slots.  To prove our result on the equilibrium of the
revenue-maximizing mechanism, we first show that a greedy algorithm
suffices to solve the revenue-maximizing linear program; we then use
this insight to prove that bidders allocated in the same block of our
mechanism have no incentive to deviate from bidding the fixed price of
that block.
\end{abstract}

\newpage

\section{Introduction}

Sponsored search is an increasingly important advertising medium,
attracting a wide variety of advertisers, large and small.  When a
user sends a query to a search engine, the advertisements are placed
into {\em slots}, usually arranged linearly down the page.  These
slots have a varying degree of exposure, often measured in terms of
the probability that the ad will be clicked; a common model is that
the higher ads tend to attract more clicks.  The problem of allocating
these slots to bidders has been addressed in various ways.  The most
common method is to allocate ads to each search independently via a
{\em generalized second price} (GSP) auction, where the ads are ranked
by (some function of) their bid, and placed into the slots in rank
order.  (See~\cite{LPSV} for a survey of this area.)  

There are several important aspects of sponsored search not captured
by the original models.  Most advertisers are interested in getting
many clicks throughout the day on a variety of searches, not just a
specific slot on a particular search query.  Also, many advertisers
have budget constraints, where they do not allow the search engine to
spend more than their budget during the day.  Finally, search engines
may have some knowledge about the distribution of queries that will
occur during the day, and so should be able to make more efficient
allocation decisions than just simple ranking.

The  {\em Offline Ad Slot Scheduling} problem is this: given a set of
bidders with bids (per click) and budgets (per day), and a set of
slots over the entire day where we know the expected number of clicks in each slot, find a
schedule that places bidders into slots.  The
schedule must not place a bidder into two different slots at the same
time.  In addition, we must find a price for each bidder that does not
exceed the bidder's budget constraint, nor their per-click bid.  (See
Section~\ref{sec:model} for a formal statement of the problem.)  

A good algorithm for this problem will have high revenue.  Also, we
would like the algorithm to be {\em truthful}; i.e., each bidder will
be incented to report her true bid and budget.  In order to prove
something like this, we need a {\em utility function} for the bidder
that captures the degree to which she is happy with her allocation.
Natural models in this context (with clicks, bids and budgets) are
{\em click-maximization}---where she wishes to maximize her number of
clicks subject to her personal bid and budget constraints, or {\em
profit-maximization}---where she wishes to maximize her profit (clicks
$\times$ profit per click).  In this paper we focus on click-maximization.\footnote{Our choice is in part motivated by the presence
of budgets, which have a natural interpretation in this application:
if an overall advertising campaign allocates a fixed portion of its
budget to online media, then the agent responsible for that budget is
incented to spend the entire budget to maximize exposure.  In
contrast, under the profit-maximizing utility, a weak motivation for budgets is a limit on liquidity.
Also, our choice of utility function is out of
analytical necessity: Borgs et al.~\cite{BCIMS} show that under some
reasonable assumptions, truthful mechanisms are impossible under a
profit-maximizing utility.}

We present an efficient mechanism for {\em Offline Ad Slot
Scheduling} and prove that it is truthful.  We also prove that the
revenue-optimal mechanism for {\em Offline Ad Slot Scheduling} is not
truthful, but has a Nash equilibrium (under the same utility model)
whose outcome is equivalent to our mechanism; this result is strong
evidence that our mechanism has desirable revenue properties.  Our
results generalize to a model where each bidder has a personal {\em
click-through-rate} that multiplies her click probability.

As far as we can tell, this is the first treatment of sponsored search
that directly incorporates both multiple positions and budget
constraints into an analysis of incentives (see
Section~\ref{sec:related} for a survey of related work).  In its full
generality, the problem of sponsored search is more complex than our
model; e.g., since the query distribution is noisy, good allocation
strategies need to be online and adaptive.  Also, our mechanism is designed for a
single query type, whereas advertisers are interested in enforcing
their budget across multiple query types.  However, the
tools used in this paper may be valuable for deriving more general
mechanisms in the future.

\subsection{Methods and Results.}
A natural mechanism for {\em Offline Ad Slot Scheduling} is the
following: find a feasible schedule and a set of prices that maximizes
revenue, subject to the bidders' constraints.  It is straightforward
to derive a linear program for this optimization problem, but
unfortunately this is not a truthful mechanism (see
Example~\ref{ex:greedy-untruthful} in Section~\ref{sec:oneslot}).
However, there is a direct truthful mechanism---the {\em price-setting} 
mechanism we present in this paper---that results in the same
outcome as an equilibrium of the revenue-maximizing mechanism. 

We derive this mechanism (and prove that it is truthful) by starting
with the single-slot case in Section~\ref{sec:oneslot}, where two
extreme cases have natural, instructive interpretations.  With only
bids (and unlimited budgets), a winner-take-all mechanism works; with
only budgets (and unlimited bids) the clicks are simply divided up 
in proportion to
budgets.  Combining these ideas in the right way results in a natural
descending-price mechanism, where the price (per click) stops at the
point where the bidders who can afford that price have enough budget
to purchase all of the clicks.

Generalizing to multiple slots requires understanding the structure of
feasible schedules, even in the special budgets-only case.  In
Section~\ref{sec:multislot} we solve the budgets-only case by
characterizing the allowable schedules in terms of the solution to a
classical {\em machine scheduling problem} (to be precise, the problem
$Q \: | \: \textit{pmtn} \: | \: C_{\max}$~\cite{graham}).  The
difficulty that arises is that the lengths of the jobs in the
scheduling problem actually depend on the price charged.  Thus, we
incorporate the scheduling algorithm into a descending-price
mechanism, where the price stops at the point where the scheduling
constraints are tight; at this point a block of slots is allocated at
a fixed uniform price (dividing the clicks equally by budget) and the
mechanism iterates.  We present the full mechanism in
Section~\ref{sec:psm} by incorporating bids analogously to the
single-slot case: the price descends until the set of bidders that can
afford that price has enough budget to make the scheduling constraints
tight.  A tricky case arises when a new bidder appears whose budget
violates the scheduling constraints; in this case the budget of this
``threshold'' bidder is reduced to make them tight again. Finally in
Section~\ref{sec:gfp} we show that the revenue-optimal mechanism has a
Nash equilibrium whose outcome is identical to our mechanism.  This
follows from the fact that if all the bidders in a block declare a bid
(roughly) equal to the price of the block, nobody has an incentive to
deviate, since every bidder is charged exactly her bid, and the clicks
are divided up equally by budget.

\subsection{Related Work.}
\label{sec:related}
There are some papers on sponsored search that analyze the {\em
generalized second-price} (GSP) auction, which is the auction
currently in use at Google and Yahoo.  The equilibria of this auction
are characterized and compared with VCG~\cite{EOS,Lahaie,AGM,Varian}.
Here the utility function is the {\em profit-maximizing}
utility where each bidder attempts to maximize her clicks $\times$
profit per click, and budget constraints are generally not treated.

Borgs et al.~\cite{BCIMS} consider the problem of budget-constrained
bidders for multiple items of a single type, with a utility function
that is profit-maximizing, modulo being under the budget (being over
the budget gives an unbounded negative utility).  They give a truthful
mechanism allocating some portion of the items that is
revenue-optimal, and prove that in their model, under reasonable
assumptions, truthful mechanisms that allocate all the units are
impossible.  Our work is different both because of the different
utility function and the generalization to multiple slots with a
scheduling constraint.  Using related methods, Mahdian et
al.~\cite{MNS} consider an online setting where an unknown number of
copies of an item arrive online, and give a truthful mechanism with a
constant competitive ratio guarantee.

There is some work on algorithms for allocating bidders with budgets to keywords
that arrive online, where the bidders place (possibly different) bids
on particular keywords~\cite{MSVV,MNS}.  The application of this work
is similar to ours, but their concern is purely online optimization;
they do not consider the game-theoretic aspects of the allocation.
Abrams et al.~\cite{AMT} derive a linear program for the offline
optimization problem of allocating bidders to queries, and handle
multiple positions by using variables for ``slates'' of bidders.  Their
LP is related to ours, but again they do not consider game-theoretic
aspects of their proposed allocations.

Bidder strategies for keyword auctions in the presence of budget
constraints have also been
considered~\cite{FMPS,RW,CDEGHKMS,BCEIJM}. Generally these papers are
not concerned with mechanism design, but there could be some
interesting relationships between the models in these papers and the
one we study here.

In our setting one is tempted to apply a {\em Fisher Market}
model: here $m$ divisible goods are available to $n$
buyers with money $B_i$, and $u_{ij}(x)$ denotes $i$'s utility of
receiving $x$ amount of good $j$.  It is known~\cite{AD,EG,DPS} that
under certain conditions a vector of prices for goods exists such that
the {\em market clears}, in that there is no surplus of goods, and all
the money is spent.  Furthermore, this price vector can be found
efficiently~\cite{DPSV}.  The natural way to apply a Fisher model to a
slot auction is to regard the slots as commodities and have the
utilities be in proportion to the number of clicks.  However this
becomes problematic because there does not seem to be a way to encode
the scheduling constraints in the Fisher model; this constraint could make an apparently
``market-clearing'' equilibrium infeasible, and indeed plays a central
role in our investigations.

\subsection{Our Setting.}
\label{sec:model}
We define the {\em Offline Ad Slot Scheduling} problem as follows.
We have $n > 1$ bidders interested in clicks.  Each bidder $i$ has a
budget $B_i$ and a maximum cost-per-click (max-cpc) $m_i$.  Given a
number of clicks $c_i$, and a price per click $p$, the utility $u_i$
of bidder $i$ is $c_i$ if both the true max-cpc and the true budget
are satisfied, and $-\infty$ otherwise.  In other words, $u_i = c_i$
if $p \leq m_i$ and $c_i p \leq B_i$; and $u_i = -\infty$ otherwise.
We have $n'$ advertising slots where slot $i$ receives $D_i$ clicks
during the time interval $[0,1]$.  We assume $\D_1 > \D_2 > \dots >
\D_{n'}$.

In a {\em schedule}, each bidder is assigned to a set of (slot, time
interval) pairs $(j, [\alpha, \beta) )$, where $j \leq n'$ and $0 \leq
\alpha < \beta \leq 1$.  A {\em feasible schedule} is one where no
more than one bidder is assigned to a slot at any given time, and no
bidder is assigned to more than one slot at any given time.
(Formally, the intervals for a particular slot do not overlap, and
the intervals for a particular bidder do not overlap.)  
A feasible schedule can be applied as follows: when a user query comes
at some time $\alpha \in [0,1]$, the schedule for that time instant is
used to populate the ad slots.  If we assume that clicks come at a
constant rate throughout the interval $[0,1]$, the number of clicks a bidder
is expected to receive from a schedule is the sum of $(\beta - \alpha)
D_j$ over all pairs $(j, [\alpha, \beta) )$ in its
schedule.\footnote{All our results generalize to the setting where
each bidder $i$ has a ``click-through rate'' $\gamma_i$ and receives $(\beta -
\alpha) \gamma_i D_j$ clicks (see Section~\ref{sec:conclusions}).  We
leave this out for clarity.}

A {\em mechanism} for {\em Offline Ad Slot Scheduling} takes as input
a declared budget $B_i$ and declared max-cpc (the ``bid'') $b_i$, and
returns a feasible schedule, as well as a price per click $p_i \leq
b_i$ for each bidder.  The schedule gives some number $c_i$ of clicks
to each bidder $i$ that must respect the budget at the given price;
i.e., we have $p_i c_i \leq B_i$.  

The {\em revenue} of a mechanism is $\sum_i p_i c_i$.  We say a
mechanism is {\em truthful} if it is a weakly dominant strategy to
declare one's true budget and max-cpc; i.e., for any particular bidder
$i$, given any set of bids and budgets declared by the other bidders,
declaring her true budget $B_i$ and max-cpc $m_i$ maximizes her
utility $u_i$.  A (pure strategy) {\em Nash equilibrium} is a set of
declared bids and budgets such that no bidder wants to change her
declaration of bid or budget, given that all other declarations stay
fixed.  An {\em $\epsilon$-Nash equilibrium} is a set of bids and
budgets where no bidder can increase her utility by more than
$\epsilon$ by changing her bid or budget.

Throughout the paper we assume some arbitrary
lexicographic ordering on the bidders, that does not necessarily match
the subscripts.  When we compare two bids $b_i$ and $b_{i'}$ we say
that $b_i \bidgt b_{i'}$ iff either $b_i > b_{i'}$, or $b_i = b_{i'}$
but $i$ occurs first lexicographically.

\section{One Slot Case}
\label{sec:oneslot}

In this section we consider the case $k=1$, where there is only one
advertising slot, with some number $\D := \D_1$ of
clicks.  We will derive a truthful mechanism for this
case by first considering the two extreme cases of infinite bids and
infinite budgets. The proofs of the theorems in this section are in Appendix~\ref{sec:proofs_single}.

Suppose all budgets $B_i = \infty$.  Then, our input amounts to bids
$b_1 \bidgt b_2 \bidgt \dots \bidgt b_n$.  Our mechanism is simply to give all the
clicks to the highest bidder.  We charge bidder 1 her full price $p_1 = b_1$.  We
claim that reporting the truth is a weakly dominant strategy for this
mechanism.  Clearly all bidders will report $b_i \leq \m_i$, since the
price is set to $b_i$ if they win.  The losing bidders cannot gain
from decreasing $b_i$.  The winning bidder can lower her price by
lowering $b_i$, but this will not gain her any more clicks, since she
is already getting all $D$ of them.

Now suppose all bids $b_i = \infty$.  In this case, our input is just
a set of budgets $B_1, \dots, B_n$, and we need to allocate $D$
clicks, with no ceiling on the per-click price.  Here we apply a
simple rule related to pricing schemes for network bandwidth
(see~\cite{Kelly,JT}): Let ${\cal B} = \sum_i B_i$.  Now to each
bidder $i$, allocate $({B_i}/{{\cal B}}) D$ clicks.  Set all prices
the same: $p_i = p = {\cal B}/D$.  The mechanism guarantees that each
bidder exactly spends her budget, thus no bidder will report $B'_i >
B_i$.  Now suppose some bidder reports $B_i' = B_i - \Delta$, for
$\Delta >0$.  Then this bidder is allocated $D({B_i - \Delta})/({{\cal B} -
\Delta})$ clicks, which is less than $D ({B_i}/{{\cal B}})$,
since $n>1$ and all $B_i > 0$.

\subsection{Greedy First-Price Mechanism.}
A natural mechanism for the general single-slot case is to solve the associated
``fractional knapsack'' problem, and charge bidders their bid; i.e.,
starting with the highest bidder, greedily add bidders to the
allocation, charging them their bid, until all the clicks are
allocated.  We refer to this as the {\em greedy first-price}
(GFP) mechanism. Though natural (and revenue-maximizing as a function of bids) this mechanism is
easily seen to be not truthful:

\begin{example}
\small
\label{ex:greedy-untruthful}
Suppose there are two bidders and $D = 120$ clicks.  
Bidder 1 has ($m_1 = \$2$, $B_1 = \$100$) and bidder 2 has ($m_2 = \$1$, $B_2 = \$50$).  
In the GFP mechanism, if both bidders tell the truth, then bidder 1
 gets 50 clicks for $\$2$ each, and 50 of the remaining 70 clicks 
go to bidder 2 for $\$1$ each.
However, if bidder 1 instead declares $b_1 = \$1 + \epsilon$, then she
gets (roughly) 100 clicks, and bidder 2 is left with (roughly) 20
clicks.
\end{example}

The problem here is that the high bidders can get away with bidding
lower, thus getting a lower price.  The difference between this and
the unlimited-budget case above is that a lower price now results in
more clicks.  It turns out that in equilibrium, this mechanism will
result in an allocation where a prefix of the top bidders are
allocated, but their prices equalize to (roughly) the lowest bid in
the prefix (as in the example above).  

\subsection{The Price-Setting Mechanism.}

An equilibrium allocation of GFP can be computed directly via the
following mechanism, which we refer to as the {\em price-setting (PS)
mechanism}.  Essentially this is a descending price mechanism: the
price stops descending when the bidders willing to pay at that price
have enough budget to purchase all the clicks.  We have to be careful
at the moment a bidder is added to the pool of the willing bidders; if
this new bidder has a large enough budget, then suddenly the willing
bidders have {\em more} than enough budget to pay for all of the
clicks.  To compensate, the mechanism decreases this ``threshold''
bidder's effective budget until the clicks are paid for exactly.  We
formalize the mechanism as follows:

\begin{mechanism}{Price-Setting (PS) Mechanism (Single Slot)}
$\bullet$ Assume wlog that $b_1 \bidgt b_2 \bidgt \dots \bidgt b_n \geq 0$.   \\
$\bullet$ Let $k$ be the first bidder such that $b_{k+1} \leq \sum_{i=1}^k B_i / D$.  Compute price $p = \min \{ \sum_{i=1}^k B_i / D, b_k \}$.\\
$\bullet$ Allocate $B_i / p$ clicks to each $i \leq k-1$.
Allocate $\hat{B}_k / p$ clicks to bidder $k$,
where $\hat{B}_k = p D - \sum_{i=1}^{k-1} B_i$.\\ \hline
\end{mechanism}

\begin{example}
\small
\label{ex:psm1}
Suppose there are three bidders with $b_1 = \$2$, $b_2 = \$1$, $b_3 =
\$0.25$ and $B_1 = \$100$, $B_2 = \$50$, $B_3 = \$80$, and $D = 300$
clicks.  
Running the PS mechanism, we get $k = 2$ since 
$B_1/D = 1/3 < b_2 = \$1$, but
$(B_1 + B_2)
/ D = \$0.50 \geq b_3 = \$0.25$.  The price is set to $\min \{ \$0.50,
\$1 \} = \$0.50$, and bidders 1 and 2 get 200 and 100 clicks at that
price, respectively.  There is no threshold bidder.
\end{example}

\begin{example}
\small
\label{ex:psm2}
Suppose now bidder 2 changes her bid to $b_2 = \$0.40$ (everything else remains the same as Example~\ref{ex:psm1}).   
We still get $k = 2$ since $B_1/D = 1/3 < b_2 = \$0.40$.  But now the
price is set to $\min \{ \$0.50, \$0.40 \} = \$0.40$, and bidders 1
and 2 get 250 and 50 clicks at that price, respectively.  Note that
bidder 2 is now a threshold bidder, does not use her entire budget, and gets fewer clicks.
\end{example}

Note that this mechanism reduces to the given mechanisms in the
special cases of infinite bids or budgets (with the proper treatment
of infinite bids/budgets).

\begin{theorem}
\label{thm:truth_single}
The price-setting mechanism (single slot) is truthful.
\end{theorem}

\subsection{Price-Setting Mechanism Computes Nash Equilibrium of GFP.}

Consider the greedy first-price auction in which the highest bidder
receives ${B_1}/{b_1}$ clicks, the second ${B_2}/{b_2}$ clicks and so
on, until the supply of $D$ clicks is exhausted.  It is immediate that
truthfully reporting budgets is a dominant strategy in this mechanism,
since when a bidder is considered, her reported budget is exhausted
as much as possible, at a fixed price.  However, reporting $b_i = m_i$
is {\em not} a dominant strategy.  Nevertheless, it turns out that GFP
has an equilibrium whose outcome is (roughly) the same as the PS
mechanism.  One cannot show that there is a plain Nash equilibrium
because of the way ties are resolved lexicographically;  
the following example illustrates why.

\begin{example}
\small
\label{ex:greedy-untruthful2}
Suppose we have the same instance as
example~\ref{ex:greedy-untruthful}: two bidders, $D = 120$ clicks,
($m_1 = \$2$, $B_1 = \$100$) and ($m_2 = \$1$, $B_2 = \$50$).  But
now suppose that bidder 2 occurs first lexicographically.
In GFP, if bidder 2 tells the truth, and bidder 1 declares $b_1 =
\$1$, then bidder 2 will get chosen first (since she is first
lexicographically), and take 50 clicks.  Bidder 2 will end up with the
remaining 70 clicks.  However, if bidder 1 instead declares $b_1 = \$1
+ \epsilon$ for some $\epsilon>0$, then she gets $100/(1 + \epsilon)$
clicks.  But this is not a best response, since she could bid $1 +
\epsilon/2$ and get slightly more clicks.
\end{example}

\noindent Thus, we prove instead that the bidders reach an $\epsilon$-Nash
equilibrium:

\begin{theorem}
\label{thm:nash_single}
Suppose the PS mechanism is run on the truthful input,
resulting in price $p$ and clicks $c_1, \dots, c_n$ for each bidder.
Then, for any $\epsilon>0$ there is a pure-strategy $\epsilon$-Nash
equilibrium of the GFP mechanism where each bidder receives $c_i \pm
\epsilon$ clicks.
\end{theorem}

\section{Multiple Slots: Bids or Budgets Only}
\label{sec:multislot}

Generalizing to multiple slots makes the scheduling constraint
nontrivial.  Now instead of splitting a pool of $D$ clicks
arbitrarily, we need to assign clicks that correspond to a feasible
schedule of bidders to slots.  The conditions under which this is
possible add a complexity that we characterize and incorporate into
our mechanism in this section.

As in the single-slot case it will be instructive to consider first
the cases of infinite bids or budgets.  Suppose all $B_i = \infty$.
In this case, the input consists of bids only $b_1 \bidgt b_2 \bidgt
\dots \bidgt b_n$.  Naturally, what we do here is rank by bid, and
allocate the slots to the bidders in that order.  Since each
budget is infinite, we can always set the prices $p_i$ equal to the
bids $b_i$. By the same logic as in the single-slot case, this is
easily seen to be truthful. In the other case, when $b_i = \infty$,
there is a lot more work to do, and we devote the remainder of the
section to this case.

Without loss of generality, we may assume the number of slots equals
the number of bids (i.e., $n' = n$); if this is not the case, then we
add dummy bidders with $B_i = b_i = 0$, or dummy slots with $D_i = 0$,
as appropriate.  We keep this assumption for the remainder of the
paper.  The proofs of the theorems in this section are in Appendix~\ref{sec:proofs_multi}.

\subsection{Assigning slots using a classical scheduling algorithm.}

  First we give an important lemma that characterizes the
conditions under which a set of bidders can be allocated to a set of
slots, which turns out to be just a restatement of a classical
result~\cite{HLS} from scheduling theory.

\begin{lemma}\label{lemma:condition}
Suppose we would like to assign an arbitrary set $\{1, \dots, k\}$ of
bidders to a set of slots $\{1, \dots, k\}$ with $D_1 > \dots >
D_k$.  Then, a click allocation $c_1 \geq ...\geq c_k$ is feasible iff
\begin{eqnarray}\label{eq:condition}
c_1 + \dots + c_\ell \leq D_1 + \dots + D_\ell \quad\textrm{ for all } \ell=1,...,k.
\end{eqnarray}
\end{lemma}
\begin{myproof}
In scheduling theory, we say a {\em job} with {\em service
requirement} $x$ is a task that needs $x / s$ units of time to
complete on a {\em machine} with {\em speed} $s$.
The question of whether there is a feasible allocation is equivalent
to the following scheduling problem: Given $k$ jobs with service
requirements $x_i = c_i$, and $k$ machines with
speeds $s_i = D_i$, is there a schedule of jobs to
machines (with preemption allowed) that completes in one unit of time?

As shown in~\cite{HLS}, the optimal schedule for this problem
(a.k.a. $Q \: | \: \textit{pmtn} \: | \: C_{\max}$) can be found
efficiently by the {\em level algorithm},\footnote{In later work,
Gonzalez and Sahni~\cite{GS} give a faster (linear-time) algorithm.}
and the schedule completes in time $\max_{\ell \leq k}
\{{\sum_{i=1}^\ell x_i}/{\sum_{i=1}^\ell s_i}\}$.  Thus, the
conditions of the lemma are exactly the conditions under which the
schedule completes in one unit of time.
\end{myproof}

\subsection{A multiple-slot budgets-only mechanism.}

Our mechanism will roughly be a
descending-price mechanism where we decrease the price until a prefix
of budgets fits tightly into a prefix of positions at that price, whereupon we
allocate that prefix, and continue to decrease the price for the
remaining bidders.

The following subroutine, which will be used
in our mechanism (and later in the general mechanism), takes a set
of budgets and determines a prefix of positions that can be packed
tightly with the largest budgets at a uniform price $p$.  The routine
ensures that all the clicks in those positions are sold at price $p$,
and all the allocated bidders spend their budget exactly.

\begin{mechanism}{Routine ``Find-Price-Block''}
Input: Set of $n$ bidders, set of $n$ slots with $D_1 > D_2 > \dots > D_n$.\\
$\bullet$ If all $D_i = 0$, assign bidders to slots arbitrarily and exit.\\
$\bullet$ Sort the bidders by budget and assume wlog that $B_1 \geq B_2 \geq ... \geq B_n$. \\
$\bullet$ Define $r_\ell = {\sum_{i=1}^{\ell} B_i}/{\sum_{i=1}^\ell D_i}$. 
Set price $p = \max_\ell r_\ell$. \\
$\bullet$
Let $\ell^*$ be the largest $\ell$ such that $r_\ell = p$.  Allocate
slots $\{1, \dots \ell^*\}$ to bidders $\{1, \dots, \ell^*\}$ at price
$p$, using all of their budgets; i.e., $c_i \defeq B_i / p$. \\ \hline
\end{mechanism}

\begin{figure}
\psfrag{#bidder#}{\bf Bidder} 
\psfrag{#budget#}{\bf Budget}

\psfrag{#b1#}{1}
\psfrag{#b2#}{2}
\psfrag{#b3#}{3}
\psfrag{#b4#}{4}

\psfrag{#B1#}{$\$80$}
\psfrag{#B2#}{$\$70$}
\psfrag{#B3#}{$\$20$}
\psfrag{#B4#}{$\$1$}

\psfrag{#35#}{$3/5$}
\psfrag{#25#}{$2/5$}
\psfrag{#2021#}{$20/21$}
\psfrag{#121#}{$1/21$}

\psfrag{#line1#}{$D_1 = 100$}
\psfrag{#line2#}{$D_2 = 50$}
\psfrag{#line3#}{$D_3 = 25$}
\psfrag{#line4#}{$D_4 = 0$}

\psfrag{#block1#}{$p_1 = \$1.00$}
\psfrag{#block2#}{$p_2 = \$0.84$}
\epsfig{file=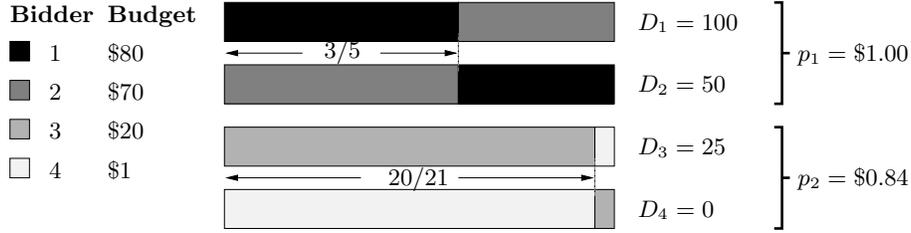,width=4.5in}
\caption{An example of the PS mechanism (multiple slots,
budgets only). We have four slots with $D_1, \dots, D_4$ clicks as
shown, and four bidders with declared budgets as shown.
The first application of Find-Price-Block computes
$r_1 = B_1 / D_1 = 80/100$, 
$r_2 = (B_1 + B_2) / (D_1 + D_2) = 150 / 150$, 
$r_3 = (B_1 + B_2 + B_3) /  (D_1 + D_2 + D_3) = 170 / 175$,
$r_4 = (B_1 + B_2 + B_3 + B_4) /  (D_1 + D_2 + D_3 + D_4) = 171 / 175$.
Since $r_2$ is largest, the top two slots make up the first price
block with a price $p_1 = r_2 = \$1$; bidder 1 gets $80$ clicks and
bidder 2 gets $70$ clicks, using the schedule as shown.
In the second price block, we get $B_3 / D_3 = 20/25$ and $(B_3 +
B_4) / (D_3 + D_4) = 21/25$.  Thus $p_2$ is set to $21/25 =
\$0.84$, bidder $3$ gets $500 / 21$ clicks and bidder $4$ gets
$25/21$ clicks, using the schedule as shown.
}
\label{ex:1}
\vspace{-.2in}
\end{figure}

\noindent Note that in the last step the allocation is always possible since for
all $\ell \leq \ell^*$, we have $p \geq r_\ell = {\sum_{i=1}^{\ell}
B_i}/{\sum_{i=1}^\ell D_i}$, which rewritten is $\sum_{i=1}^\ell c_i
\leq \sum_{i=1}^\ell D_i$, and so we can apply
Lemma~\ref{lemma:condition}.
Now we are ready to give the mechanism in terms of this subroutine; an example run is 
shown in Figure~\ref{ex:1}.

\begin{mechanism}{Price-Setting Mechanism (Multiple Slots, Budgets Only)}
$\bullet$ Run ``Find-Price-Block'' on bidders $1, \dots, n$, and slots $1,
\dots, n$.  This gives an allocation of $\ell^*$ bidders to the first $\ell^*$ slots.  \\
$\bullet$ Repeat on the remaining bidders and slots until all slots are allocated. \\ \hline
\end{mechanism}

\noindent 
Let $p_1, p_2, \dots$ be the prices used for each successive block
assigned by the algorithm.  We claim that $p_1 > p_2 > \dots$; to see
this, note then when $p_1$ is set, we have $p_1 = r_k$ and $p_1 >
r_\ell$ for all $\ell > k$, where $k$ is the last bidder in the block.
Thus for all $\ell > k$, we have $p_1 \sum_{j \leq \ell} D_j > \sum_{i
\leq \ell} B_j$, which gives $p_1 \sum_{k < j \leq \ell} D_j > \sum_{k
< i \leq \ell} B_j$ using $p_1 = r_k$.  This implies that when we
apply Find-Price-Block the second time, we get $r'_\ell =
{\sum_{k < i \leq \ell} B_j}/{\sum_{k < j \leq \ell} D_j} < p_1$,
and so $p_2 < p_1$.  This argument applies to successive blocks to
give $p_1 > p_2 > \dots$.

\begin{theorem}\label{thm:monotonicity}
The price-setting mechanism (multiple slots, budgets only) is truthful. 
\end{theorem}

\section{Main Results}
\label{sec:psm}

In this section we give our main results, presenting our price-setting
mechanism in the general case, building on the ideas in the previous
two sections.  We begin in Section~\ref{sec:mech_gen} by stating the
mechanism and showing some examples, then proving that the mechanism is
truthful.  In Section~\ref{sec:gfp} we
analyze the revenue-optimal schedule, and show that it can be
computed with a generalization of the {\em greedy first-price (GFP)}
mechanism.  We then show that GFP has an $\epsilon$-Nash equilibrium
whose outcome is identical to the general PS mechanism.
The proofs of the theorems in this section are in Appendix~\ref{sec:proofs_gen}.

\subsection{The Price-Setting Mechanism (General Case).}
\label{sec:mech_gen}

The generalization of the PS mechanism combines the ideas
from the bids-and-budgets version of the single slot mechanism with
the budgets-only version of the multiple-slot mechanism.  As our price
descends, we maintain a set of ``active'' bidders with bids at or
above this price, as in the single-slot mechanism.  These active
bidders are kept ranked by {\em budget}, and when the price reaches
the point where a prefix of bidders fits into a prefix of slots (as in
the budgets-only mechanism) we allocate them and repeat.  As in the
single-slot case, we heave to be careful when a bidder enters the
active set and suddenly causes an over-fit; in this case we again reduce the
budget of this ``threshold'' bidder until it fits.  We formalize this as follows:

\begin{mechanism}{Price-Setting Mechanism (General Case)}
(i) Assume wlog that $b_1 \bidgt b_2 \bidgt \dots \bidgt b_n = 0$.  \\
(ii) Let $k$ be the first bidder such that running Find-Price-Block on
bidders $1, \dots, k$ would result in a price $p \geq b_{k+1}$.\\
(iii) Reduce $B_k$ until running Find-Price-Block on bidders $1,
\dots, k$ would result in a price $p \leq b_k$.  Apply this
allocation, which for some $\ell^* \leq k$ gives the first $\ell^*$
slots to the $\ell^*$ bidders among $1 \dots k$ with the largest budgets.\\ 
(iv) Repeat on the remaining bidders and slots. \\ \hline
\end{mechanism}

\begin{figure}
\psfrag{#bidder#}{\bf Bidder} 
\psfrag{#budget#}{\bf Budget}
\psfrag{#bid#}{\bf Bid}

\psfrag{#b1#}{1}
\psfrag{#b2#}{2}
\psfrag{#b3#}{3}
\psfrag{#b4#}{4}

\psfrag{#bid1#}{$\$3$}
\psfrag{#bid2#}{$\$0.75$}
\psfrag{#bid3#}{$\$1$}
\psfrag{#bid4#}{$\$0.50$}

\psfrag{#B1#}{$\$80$}
\psfrag{#B2#}{$\$70$}
\psfrag{#B3#}{$\$20$}
\psfrag{#B4#}{$\$1$}

\psfrag{#s1#}{$29/45$}
\psfrag{#s2#}{$16/45$}

\psfrag{#line1#}{$D_1 = 100$}
\psfrag{#line2#}{$D_2 = 50$}
\psfrag{#line3#}{$D_3 = 25$}
\psfrag{#line4#}{$D_4 = 0$}

\psfrag{#block2#}{$p_1 = \$0.80$}
\psfrag{#block1#}{$p_2 = \$0.75$}
\psfrag{#block3#}{$p_3 = \$0$}
\epsfig{file=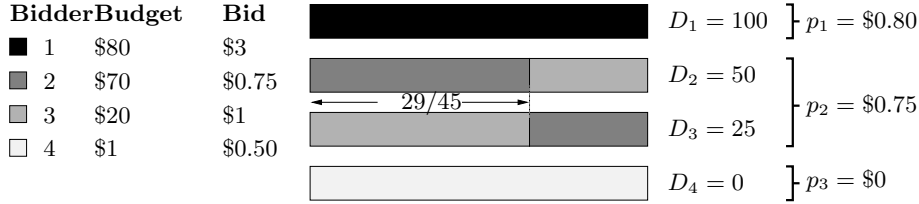,width=4.5in}
\caption{
Consider the same bidders and slots as in Figure~\ref{ex:1}, but now add bids
as shown.
Running Find-Price-Block on only bidder 1 gives a price of $r_1 =
80/100$, which is less than the next bid of $\$1$.  So, we run
Find-Price-Block on bidders 1 and 3 (the next-highest bid), giving
$r_1 = 80/100$ and $r_2 = 100/150$.  We still get a price of $\$0.80$,
but now this is more than the next-highest bid of $\$0.75$, so we
allocate the first bidder to the first slot at a price of $\$0.80$.
We are left with bidders 2-4 and slots 2-4.  With just bidder 3 (the
highest bidder) and slot 2, we get a price $p = 20/50$ which is less
than the next-highest bid of $\$0.75$, so we consider bidders 2 and 3
on slots 2 and 3.  This gives a price of $\max \{ 70/50, 90/75 \} =
\$1.40$, which is more than $\$0.50$.  Since this is also more than
$\$0.75$, we must lower $B_2$ until the price is exactly $\$0.75$,
which makes $B'_2 = \$36.25$.  With this setting of $B'_2$,
Find-Price-Block allocates bidders 2 and 3 to slots 2 and 3, giving
$75(36.25/56.25)$ and $75(20/56.25)$ clicks respectively, at a price
of $\$0.75$ per click.  Bidder 4 is allocated to slot 4, receiving
zero clicks.  }
\label{ex:2}
\vspace{-.2in}
\end{figure}

\noindent An example run of this mechanism is shown in Figure~\ref{ex:2}.  
Since the PS mechanism sets prices per slot, 
it is natural to ask if these prices constitute some sort of
``market-clearing'' equilibrium in the spirit of a Fisher market.  The
quick answer is no: since the price per click increases for higher
slots, and each bidder values clicks at each slot equally, then
bidders will always prefer the bottom slot.
Note that by the same logic as the budgets-only mechanism, the prices $p_1, p_2,
\dots$ for each price block strictly decrease.

\subsubsection{Efficiency.}
So far we have been largely ignoring the efficiency of computing the
allocation in the PS mechanism.  It is immediately clear
that the general PS mechanism can be executed in time
polynomial in $n$ and $\log(1/\epsilon)$ to some precision $\epsilon$
using binary search and linear programming.  

In fact, a purely combinatorial $O(n^2)$ time algorithm is possible.
As bidders get added in step (ii), maintaining a sorted list of
bidders and budgets can be done in time $O(n \log n)$.  Thus it
remains to show that running Find-Price-Block (and computing the
reduced budget) can be done in $O(n)$ time given these sorted lists.  In
Find-Price-Block, computing the ratios $r_\ell$ can be done in linear
time.  Finding the allocation from Lemma~\ref{lemma:condition} can
also be done in linear time using the Gonzalez-Sahni
algorithm~\cite{GS} for scheduling related parallel machines (in fact
the total time for scheduling can be made $O(n)$ since each slot is
scheduled only once).  Finally, computing the reduced budget is a
simple calculation on each relevant ratio $r_\ell$, also doable in
linear time.  We suspect that there is a $O(n \cdot \mathrm{polylog}(n))$ algorithm
using a more elaborate data structure; we leave this open.

\begin{theorem}
\label{thm:truth_gen}
The price-setting mechanism (general case) is truthful.
\end{theorem}

\subsection{Greedy First-Price Mechanism for Multiple Slots.}
\label{sec:gfp}

In the general case, as in the single-slot case, there is a natural
{\em greedy first-price} mechanism when the bidding language includes
both bids and budgets: Order the bidders by bid $b_1 \bidgt b_2 \bidgt \dots \bidgt
b_n$. Starting from the highest bidder, for each bidder $i$ compute
the maximum possible number of clicks $c_i$ that one could allocate to
bidder $i$ at price $b_i$, given the budget constraint $B_i$ and the
commitments to previous bidders $c_1, \dots, c_{i-1}$.  This reduces
to the ``fractional knapsack'' problem in the single-slot case, and so
one would hope that it maximizes revenue for the given bids and
budgets, as in the single-slot case.  This is not immediately clear,
but does turn out to be true, as we will prove in this section.

As in the single-slot case, the greedy mechanism is not a truthful
mechanism.  However, we show that it does have a pure-strategy
equilibrium, and that equilibrium has prices and allocation equivalent
to the price setting mechanism.

\subsubsection{Greedy is Revenue-Maximizing.} 
\label{sec:equivalence}

Consider a revenue-maximizing schedule that respects both bids and
budgets.  In this allocation, we can assume wlog that each bidder $i$
is charged exactly $b_i$ per click, since otherwise the allocation can
increase the price for bidder $i$, reduce $c_i$ and remain feasible.
Thus, by Lemma~\ref{lemma:condition}, we can find a revenue-maximizing
schedule $\carr^* = (c^*_1, \dots, c^*_n)$ by maximizing
$\sum_{i}{b_i c_i}$ subject to $c_i \leq B_i / b_i$
and
$
c_1 + \dots + c_\ell \leq D_1 + \dots + D_\ell$ for all $\ell=1,...,n.
$

\begin{theorem}\label{thm:equivalence}
The greedy first-price auction gives a revenue-maximizing schedule.
\end{theorem}

\subsubsection{Price-Setting Mechanism is a Nash Equilibrium of the Greedy First Price Mechanism.}
\label{sec:greedy-nash}

We note that truthfully reporting one's budget is a weakly dominant
strategy in GFP, since when a bidder is considered for allocation,
their budget is exhausted at a fixed price, subject to a cap on the
number of clicks they can get.  Reporting one's bid truthfully is not a
dominant strategy, but we can still show that there is an
$\epsilon$-Nash equilibrium whose outcome is arbitrarily close to the
PS mechanism.

\begin{theorem}
\label{thm:nash}
Suppose the PS mechanism is run on the truthful input,
resulting in clicks $c_1, \dots, c_n$ for each bidder.  Then, for any
$\epsilon>0$ there is a pure-strategy $\epsilon$-Nash equilibrium of the GFP
mechanism where each bidder receives $c_i \pm \epsilon$ clicks.
\end{theorem}

\section{Conclusions}
\label{sec:conclusions}

In this paper we have given a truthful mechanism for assigning bidders
to click-generating slots that respects budget and per-click price
constraints.  The mechanism also respects a scheduling constraint on
the slots, using a classical result from scheduling theory to
characterize (and compute) the possible allocations. We have also
proved that the revenue-maximizing mechanism has an $\epsilon$-Nash
equilibrium whose outcome is arbitrarily close to our mechanism.  This
final result in some way suggests that our mechanism is the right one
for this model.  It would interesting to make this more formal; we
conjecture that a general truthful mechanism cannot do better in terms
of revenue.

\subsection{Extensions.} There are several natural generalizations of the 
{\em Online Ad Slot Scheduling} problem where it would be interesting
to extend our results or apply the knowledge gained in this paper.  We
mention a few here.

\paragraph{Click-through rates.} In sponsored search (e.g.~\cite{EOS}) it is common for each bidder to
have a personal click-through-rate $\gamma_i$; in our model this would
mean that a bidder $i$ assigned to slot $j$ for a time period of
length $\alpha$ would receive $\alpha \gamma_i D_j$ clicks.  All our
results can be generalized to this setting by simply scaling the bids
using $b'_i = b_i \gamma_i$.  However, our mechanism in this case does
not necessarily prefer more {\em efficient} solutions; i.e., ones that generate
more overall clicks.  It would be interesting to analyze a possible
tradeoff between efficiency and revenue in this setting.

\paragraph{Multiple Keywords.} To model multiple keywords in our model, we
could say that each query $q$ had its own set of click totals $D_{q,
1} \dots D_{q, n}$, and each bidder is interested in a subset of
queries.  The greedy first-price mechanism is easily generalized to
this case: maximally allocate clicks to bidders in order of their bid
$b_i$ (at price $b_i$) while respecting the budgets, the query
preferences, and the click commitments to previous bidders.  It would
not be surprising if there was an equilibrium of this extension of 
the greedy mechanism that could
be computed directly with a generalization of the PS
mechanism.

\paragraph{Online queries, uncertain supply.} In sponsored search,
allocations must be made online in response to user queries, and some
of the previous literature has focused on this aspect of the problem
(e.g., \cite{MSVV,MNS}). Perhaps the ideas in this paper could be
used to help make online allocation decisions using (unreliable)
estimates of the supply, a setting considered in~\cite{MNS}, with game-theoretic
considerations.

\bigskip

\noindent {\bf Acknowledgments.}
We thank Cliff Stein and Yishay Mansour for helpful discussions.

\bibliographystyle{plain}
\bibliography{throttling}

\appendix

\section{Proofs for Section~\ref{sec:oneslot}}
\label{sec:proofs_single}

\begin{proofof}{Theorem~\ref{thm:truth_single}}
For the purposes of this proof, let bidders $\{1, \dots, n\}$ be such
that $b_1 \bidgt \dots \bidgt b_n = 0$, and consider a new bidder (call her Alice) with true max-cpc
$m$ and true budget $B^*$.  

We first show that reporting the true budget is a weakly dominant
strategy for Alice, for any fixed bid $b > 0$.  Let $\ell$ be
the first bidder with $b \bidgt b_\ell$, so $b_1 \bidgt \dots \bidgt b_{\ell-1}
\bidgt b \bidgt b_\ell \bidgt \dots \bidgt b_n$. Let ${\cal B} = \sum_{i=1}^{\ell-1} B_i$.  If
${\cal B} \geq b D$ then the mechanism will not allocate any clicks to
Alice, regardless of the reported budget, since the price
will stop before reaching $b$.  
If ${\cal B} < b D$, we will argue that Alice's clicks $c$ are non-increasing in $B$.  Define $\hat{B} = b D - {\cal
B} > 0$. 
\begin{itemize}
\item
If Alice declares $B \in [\hat{B}, \infty]$, then the price will stop at $b$.  She will spend $\hat{B}$ and
receive $c = \hat{B}/b$ clicks.  
\item
If Alice declares $B \in [0,
\hat{B})$, then the price will be lower than $b$, and she will spend
all of her budget.  Her final number of clicks will be $c = ({B}/({B + {\cal B} + R})) D$, where $R$ is the total spend of bidders $\{\ell, \dots, n\}$.  Since $R$ is
non-increasing in $B$, we can conclude that $c$ is non-decreasing in
$B$. 
\end{itemize}
Putting together these intervals, we see that $c$ is
non-decreasing in $B$ overall, and since Alice's
total spend is $\min\{B, \hat{B}\}$, we may conclude that it is weakly
dominant to declare $B = B^*$.

\medskip

It remains to show that it is weakly dominant for Alice to
declare a bid $b = m$, given that she declares a budget $B = B^*$.  Let
$R(b)$ be the total spend of bidders $\{1, \dots, n\}$ given that Alice declares $b$.  Note that $R(b)$ is non-increasing in $b$.
Let $p_1$ be the price that would result if $b = \infty$, and let
$p_2$ be the price that would result if $b = 0$. Note that $p_2 \leq p_1$.

\begin{itemize}
\item
If $b \in [0, p_2)$ then the price stops at $p_2$ and Alice receives zero clicks.
\item
If $b \in (p_1, \infty]$, then the price stops at $p_1$, and Alice receives $B/p_1$ clicks.
\item
If $b \in [p_2, p_1]$, then the price stops at $b$.  To see this, note
that if Alice had bid zero, then the price would have gone down
to $p_2$, so it certainly stops at $b$ or lower.  But at price $b$, the
set of bidders that can afford this price consists of at least all the bidders
 that could afford price $p_1$, and so we must have $B + \sum_{i:b_i \bidgt
b} B_i \geq B + \sum_{i:b_i \geq p_1} B_i \geq p_1 D \geq b D$.  
Alice thus receives 
\begin{equation}
\label{eq:casemidbid}
\max \bigg \{ 0, D - \bigg (\sum_{i:b_i \bidgt b} B_i/b \bigg ) \bigg \}
\end{equation}
clicks, and we may conclude that in this interval, clicks are
non-decreasing with $b$.
\end{itemize}
Note that in the expression~\eqref{eq:casemidbid}, plugging in $p_1$
for $b$ yields $c = B/p_1$.  Thus we have that in the interval $[p_2,
\infty]$, clicks are non-decreasing with $b$, and the price is always
$\min \{b, p_1\}$.  We conclude that bidding $b = m$ is a weakly
dominant strategy.
\end{proofof}

\bigskip

\begin{proofof}{Theorem~\ref{thm:nash_single}}
We will show that for sufficiently small $\epsilon'>0$, if each bidder
truthfully reports her budget and bids $b_i = \min \{m_i, p +
\epsilon'\}$ in the GFP mechanism, then the conditions in the theorem
hold.

There are two ways that the PS mechanism (under truthful input) can
reach its last allocated bidder $k$ and final price $p$: if $m_k > p
\geq m_{k+1}$ and then $pD = \sum_{i=1}^k B_i$ (no threshold bidder),
or if $p = m_k$ ($k$ is a threshold bidder).

In the first case, we have that bidders $i \leq k$ all have $m_i >
p$. Thus in the supposed equilibrium of GFP, all these bidders are
bidding $p+\epsilon'$, and all bidders $i > k$ are bidding $m_i \leq
p$.  Therefore in GFP, each $i \leq k$ will receive $B_i /
(p+\epsilon')$ clicks, and the total number of clicks allocated by GFP
to bidders $1 \dots k$ is $\sum_{i \leq k} B_i / (p+\epsilon') =
(\frac{p}{p+\epsilon'}) D$.  The remaining $D' = (1 -
\frac{p}{p+\epsilon}) D$ clicks, are allocated to bidders $i > k$.
Bidders $1 \dots k$ lose clicks by increasing their bid, and can gain at
most $D'$ clicks by lowering their bid.  Bidders $i > k$ will never
raise their bid (since they are bidding $m_i$), and cannot gain more
clicks by lowering their bid.  Since $D'$ can be made arbitrarily
small, we have an $\epsilon$-Nash equilibrium.

In the second case, $p = m_k$.  Let $k' < k$ be the last bidder
bidding more than $p$.  In the supposed GFP equilibrium, bidders $1
\dots k'$ are bidding $p+\epsilon'$, and bidders $(k'+1, \dots, k)$
are bidding $m_k = p$.  Thus GFP allocates $B_i / (p+\epsilon')$
clicks to bidders $1 \dots k'$, $B_i / p$ clicks to bidders $(k'+1,
\dots, k-1)$ (if any such bidders exist) and the remaining clicks to
bidder $k$.  As in the previous case, no bidder can gain from raising
her bid, the number of clicks that a bidder $i \leq k'$ can gain from
lowering her bid can be made arbitrarily small, and no other bidder
can gain from lowering her bid.
\end{proofof}

\section{Proofs for Section~\ref{sec:multislot}}
\label{sec:proofs_multi}

\begin{lemma}\label{lemma:equal-budgets}
In Find-Price-Block, if $B_i = B_{i+1}$, then $i$ cannot be the last slot of the computed price block. 
\end{lemma}
\begin{myproof}
Suppose the contrary, namely that $i$ is the last slot of the first price block
and $(i+1)$ is the first slot in the second price block. 
Denote $B = B_1 + ... + B_{i-1}$ and $D = D_1 + ... + D_{i-1}$.
Then the price of the first price block satisfies 
(1) $p_1 = \frac{B+B_i}{D+D_i} \geq \frac{B}{D}$ and
(2) $p_1 = \frac{B+B_i}{D+D_i} > \frac{B+B_{i}+B_{i+1}}{D+D_i+D_{i+1}}$.
The first condition is equivalent to 
$\frac{B_i}{D_i} \geq \frac{B+B_i}{D+D_i}$,
and the second condition is equivalent to 
$\frac{B+B_i}{D+D_i} > \frac{B_{i+1}}{D_{i+1}}$. 
The latter two inequalities imply $\frac{B_i}{D_i} > \frac{B_{i+1}}{D_{i+1}}$, 
which is a contradiction to the fact that $B_i = B_{i+1}$ and $D_i > D_{i+1}$.
\end{myproof}

\bigskip

\begin{proofsketchof}{Theorem~\ref{thm:monotonicity}}
Suppose bidders $1, \dots, n$ declare budgets $B_1 \geq \dots \geq
B_n$, and Alice declares budget $B$.  Let $\ell_B$ be the rank of
Alice by budget (and lexicographic order in case of ties) if she bids
$B$.  We will prove that the number of clicks Alice receives is
non-increasing as she lowers her declared budget $B$, which immediately implies
that truthful reporting of budgets is weakly dominant in the
PS mechanism.

Let $r^B_j$ be the ratio $r_j$ assuming Alice bids $B$; 
so $r_k^B = ({B + \sum_{i=1}^{k-1} B_i})/{\sum_{i=1}^k D_i}$ if $\ell_B \leq k$, 
and $r_k^B = {\sum_{i=1}^{k} B_i}/{\sum_{i=1}^k D_i}$ otherwise.
For a declared budget $B$, let $k_B$ be the last slot in the first
price block chosen by the mechanism.  So, $k_B = \argmax_k r_k^B$ (if
there are multiple maxima, then $k_B$ is the largest
lexicographically).

For sufficiently large $B > B_1$, we get that $r^B_1 > r^B_k$ for all $k$ and so $k_B = 1$.
For any such $B$ Alice receives $D_1$ clicks, the most possible.  Now
as we lower $B$, two significant events could occur; we could drop
to another bidder's budget $B_i$, or we could have a change in
$k_B$, thus changing the set of bidders in the first
block.  If neither of these events occur, then Alice remains in the first price block, but gets a smaller share of the clicks.  Thus it
remains to cover these two events.

If $B = B_i$ for some $i$, then note that by
Lemma~\ref{lemma:equal-budgets}, Alice cannot be the last bidder in
the block, so $i$ is in the same block as Alice.  Therefore we may
exchange the roles of Alice and bidder $i$ lexicographically (i.e.,
increase Alice's rank by one) and nothing changes.

Now suppose $B$ reaches a point where $r_k$ changes because $\argmax_k
r_k^B$ changes from $k_B$ to $k'$. We use $k^* = k_B$ for the remainder of the proof for ease of notation.
At the bid $B$ we have $r^B_{k^*} = r^B_{k'}$.
We claim that either $k' > k^*$ or $k' < \ell_B$.  To see this note
that for any $k$ between $\ell_B$ and $k^*$ we have that $r_k^B$
decreases at a rate of $1/(\sum_{i=1}^k D_i)$, which is faster than
the rate of the highest ratio $r_{k^*}^B$.

If $k' > k^*$ then Alice remains in the first block, but it expands
from ending at $k^*$ to ending at $k'$.  Both before and after the
change in $r_k$, Alice is spending her entire budget at price
$r^B_{k^*} = r^B_{k'}$, so her clicks remain the same.

If $k' < \ell_B$ then Alice would remain in a block ending at slot
$k^*$, since $r_{k^*}^B$ remains maximum among
$r^B_{\ell_B},...,r^B_n$ (by the same reasoning about ``rate'' as
above).  Since $r^B_{k^*} = r^B_{k'}$ we have that the price of
Alice's block and the first block will be the same.  Since Alice is
spending her entire budget before and after the change in $r_k$ at the
same price, her clicks remain the same.  As we continue to decrease
$B$ beyond this point, we simply remove the bidders and slots from the
first price block, and imagine that we are again in the first price
block of a reduced instance.
\end{proofsketchof}

\section{Proofs for Section~\ref{sec:psm}}
\label{sec:proofs_gen}

\begin{proofsketchof}{Theorem~\ref{thm:truth_gen}}
We split the proof into two lemmas, showing that clicks are
non-decreasing in both bids and budgets.  This immediately implies the
theorem.  First we need a small observation about Find-Price-Block:

\begin{lemma}\label{lemma:new-bidder}
Suppose Find-Price-Block is run on a set of budgets $B_1 \geq \dots
\geq B_n$ and produces a block $1, \dots, \ell^*$ with price $p$.
Then if a bidder is added to the set with budget $B$, and
Find-Price-Block still produces price $p$, we must have that $B \leq
B_{\ell^*}$.
\end{lemma}

\begin{myproof}
Suppose not.  Then $B > B_{\ell*}$ and we have that $({B +
\sum_{i=1}^{\ell^*-1} B_i})/{\sum_{i=1}^{\ell^*} D_i} \leq p$.  This
contradicts $p={\sum_{i=1}^{\ell^*} B_i}/{\sum_{i=1}^{\ell^*}
D_i}$, since $B > B_{\ell*}$.
\end{myproof}

\begin{lemma}
\label{lemma:nondec_budget}
The number of clicks a bidder is allocated is non-decreasing in her
declared budget. 
\end{lemma}
\begin{proofsketch}
Let bidders $\{1, \dots, n\}$ be such that $b_1 \bidgt \dots \bidgt
b_n$, and consider a new bidder Alice with bid $b_{\ell-1} \bidgt b
\bidgt b_\ell$.  We will argue that the number of clicks that Alice
receives is non-increasing as she reduces her declared budget $B$.

Suppose Alice declares $B = \infty$ and let $\hat{B}$ be the amount
she would spend (Alice would always be a threshold bidder if she
declared $B = \infty$).  Any declared budget $B \in [\hat{B}, \infty]$
would result in the same number of clicks, because $B$ is reduced by
the mechanism in step (iii) to $\hat{B}$.

Now as $B$ decreases from $\hat{B}$, two different events could occur:
(a) Alice's price block threshold $\ell^*$ could change (because
Find-Price-Block outputs a different $\ell^*$) or (b) the lowest
bidder $k$ could change (because running Find-Price-Block on $1,
\dots, k$ gave a price less than $b_{k+1}$).  For event (a), and
between these events, the arguments from
Theorem~\ref{thm:monotonicity} imply that Alice's clicks are
non-increasing.

For event (b), when the price of the Alice's block is exactly
$b_{k+1}$, if bidder $k+1$ is added, the resulting price output by
Find-Price-Block in step (ii) is still at least $b_{k+1}$, since
adding a bidder cannot reduce the price.  Also Lemmas~\ref{lemma:new-bidder}
and~\ref{lemma:equal-budgets} 
together imply that Alice is still in the price block chosen in step (iii).
Thus Alice's clicks do not increase.
\end{proofsketch}

\begin{lemma}
\label{lemma:nondec_bid}
The number of clicks a bidder is allocated is non-decreasing in her 
declared bid. 
\end{lemma}

\begin{proofsketch}
For the purposes of this proof, let bidders $\{1, \dots, n\}$ be such
that $b_1 \bidgt \dots \bidgt b_n$, and consider a new bidder (call
her Alice) with declared budget $B$.  We will argue that the number of clicks
that Alice receives in non-increasing with her declared bid $b$.

Let $p_1$ be the price that Alice would pay if $b = \infty$, and
suppose Alice is in the $j$th price block when she bids $\infty$.
Note that for any bid $b \in (p_1, \infty]$, Alice is still in the
$j$th price block and receives the same number of clicks ($B/p_1$).
Let $p_2$ be the minimum bid required to keep Alice in the $j$th price
block.

We claim that if $b \in [p_2, p_1]$, the price will always be exactly
$b$: no allocation is made until Alice is considered in step (ii), and
when she's considered, Find-Price-Block returns a price $p \geq p_1$,
since the set of bidders considered contains all the bidders who
produced price $p_1$.  Thus Alice is a threshold bidder, and in step $(iii)$ Alice's budget is reduced
so that the price is exactly $b$.

Let $k_b$ be the number of bidders with bid $b_i \bidgt b$.  Let
$B^b_i$ be the $i$th largest budget among bidders with bid $b_i \bidgt
b$.  We claim that if $b \in [p_2, p_1]$, we have
$
{\sum_{i=1}^\ell B^b_i}/{\sum_{i=1}^\ell D_i} < b
$ for all $\ell \leq k_b$, since otherwise Alice would not be in the $j$th block.

Let $\hat{B}_b$ be Alice's reduced budget when she bids $b \in [p_2,
p_1]$, and let $c_b = \hat{B}_b / b$ denote the number of clicks she receives.
To satisfy the price being at most $b$ in step (iii), we must have that for all $\ell \leq k_b$, 
$
\hat{B}_b \leq B^b_\ell + \Delta
$, 
where $\Delta > 0$ satisfies
$
({\Delta + \sum_{i=1}^\ell B^b_i})/{\sum_{i=1}^\ell D_i} = b
$.  
In addition, we must have 
$
({B_b + \sum_{i=1}^{k_b} B^b_i})/{\sum_{i=1}^{k_b+1} D_i} \leq b
$.
Putting these constraints together we get
$
\hat{B}_b 
= \min_{\ell \leq k_b+1} \{b \sum_{i=1}^{\ell} D_i - \sum_{i=1}^{\ell-1} B^b_i  \}
$ 
and so 
$$
c_b = \hat{B}_b / b 
= \min_{\ell \leq k_b+1} \left \{\sum_{i=1}^{\ell} D_i - \frac 1 b \sum_{i=1}^{\ell-1} B^b_i \right \}.
$$ 
As $b$ decreases, if the set of bidders with bids $\bidgt b$
doesn't change, then the $B^b_i$s don't change, and so this expression
implies that $c_b$ also decreases.  If $b$ decreases to the point
where $b' \bidgt b$ for some new bidder $b'$, then we claim that $c_b$
also cannot increase.  To see this note that for all $\ell$, the
expression $\sum_{i=1}^{\ell - 1} B^b_i$ can only increase or stay the
same if a new bidder is added.  We conclude that $c_b$ is
non-increasing in the interval $b \in [p_2, p_1]$.

\medskip

When $b$ decreases to $p_2$, we transition from Alice being in the
$j$th price block to the $j+1$st price block. As in
Theorem~\ref{thm:monotonicity}, at the point of transition the $j$th
price block will have the same price as the $j+1$st price block, and
in both scenarios Alice spends exactly $\hat{B}_{p_2}$.  Thus her clicks do
not change.  
We can iterate these arguments for the $j+1$st price
block, and so the theorem is proven.
\end{proofsketch}

\medskip

Lemmas~\ref{lemma:nondec_budget} and~\ref{lemma:nondec_bid} immediately
imply Theorem~\ref{thm:truth_gen}.
\end{proofsketchof}

\bigskip

\begin{proofof}{Theorem~\ref{thm:equivalence}}
Note that an equivalent statement of the constraint 
$c_1 + \dots + c_\ell \leq D_1 + \dots + D_\ell$ for all $\ell=1,...,n.$
is:
\begin{eqnarray}\label{eq:condition3}
\sum_{i\in S}{c_i'} \leq D_1+...+D_{|S|} 
    \quad\textrm{ for all subsets } S\subseteq\{1,...,n\}.
\end{eqnarray}

Suppose bids are $b_1 \bidgt b_2 \bidgt ... \bidgt b_n$ and the
corresponding clicks given to bidders in the greedy
allocation are $\carr = (c_1,...,c_n)$.  Let $\carr^{*} =
(c_1^{*},...,c_n^{*})$ be the revenue-maximizing solution with the
{\em closest prefix} to $\carr$, meaning that the first $i$ such that
$c_i \neq c^*_i$ is maximized, and modulo that, $c_i - c^*_i$ is
minimized.

We shall prove that the greedy $\carr$ gives a revenue-maximizing schedule.  
Suppose the contrary
and let $i$ be the first index on which $\carr$ differs from $\carr^*$.  Note that 
$c_i > c_i^*$ (by the definition of greedy, $c_i$ is the maximum possible 
given $c_1,...,c_{i-1}$).
Let $\cmax = \max\{c^*_{i+1},...,c^*_n\}$.
Let $J = \{j > i : c^*_j = \cmax \}$.
Consider an arbitrary tight constraint on $\carr^*$ of the
form~(\ref{eq:condition3}), defined by the set $S$.  We claim that if
$i \in S$, then all $j \in J$ are also in $S$.  

\medskip
\begin{proofof}{claim}
Suppose
the contrary, namely that $i \in S$ and $j \notin S$ for some $j \in J$.
Applying~\eqref{eq:condition3}, we get 
\begin{eqnarray}
\label{eq:tightconstraint}
\sum_{\ell\in S} c^*_\ell = \sum_{\ell \leq |S|} {D_\ell}.
\end{eqnarray}
One of the bidders in $S$ must have index $m > i$, 
otherwise~\eqref{eq:condition3} would be violated for $c$ and $S$ by
$\sum_{\ell\in S \subseteq \{1,...,i\}} c_\ell > 
  \sum_{\ell\in S \subseteq \{1,...,i\}} c^*_\ell = 
\sum_{\ell\leq |S|} {D_\ell}.$
If $m \notin J$, then
we would violate~\eqref{eq:condition3} for the set $S' = S \cup \{j\} \backslash \{m\}$:
$
\sum_{\ell \in S'} c^*_\ell > \sum_{\ell \in S} c^*_\ell = \sum_{\ell\leq |S| = |S'|} D_\ell
$.
Therefore $m \in J$.  

Now by the feasibility of $\carr^*$ and the fact that $j \notin S$, we also have
$
c^*_j + \sum_{\ell\in S} c^*_\ell \leq D_{|S|+1} + \sum_{\ell \leq |S|} {D_\ell}
$
which implies, together with~\eqref{eq:tightconstraint}, that $c^*_j \leq D_{|S|+1}$.
Again by feasibility, we also have 
$
\sum_{\ell\in S \setminus m} c^*_\ell \leq \sum_{\ell \leq |S| - 1} {D_\ell}
$
and this, together with~\eqref{eq:tightconstraint}, gives $c^*_m \geq D_{|S|}$.
Putting these last two observations together yields $D_{|S|} \leq c_m
= \cmax = c^*_j \leq D_{|S|+1}$.  Unless $c_m = \cmax = c^*_j = 0$,
this violates the distinctness of the non-zero $D_j$'s.  But if $\cmax
= 0$, it means that all $c_\ell$ for $\ell > i$ have $c_\ell = 0$,
which means that $\carr$ gives strictly more clicks than
 $\carr^*$, a contradiction.
\end{proofof}

\medskip

Let $j$ be an arbitrary member of $J$.  By the claim, there is an
$\epsilon>0$ such that if we set $\carr' = \carr^*$ except $c'_i =
c^*_i + \epsilon$ and $c'_j = c^*_j - \epsilon$, we get a feasible
allocation $\carr'$, since $j$ appears in every tight constraint in
which $i$ appears.  This allocation has revenue at least that of
$\carr^*$, since $b_i \geq b_j$.  But, it has a closer prefix to
$\carr$ than $\carr^*$, a contradiction.
\end{proofof}

\bigskip

\begin{proofsketchof}{Theorem~\ref{thm:nash}}
We will abuse notation and let $\epsilon'$ denote any positive
quantity that can be made arbitrarily close to zero.  When the
PS mechanism is run on the truthful input, let $p_1 > p_2 >
\dots$ denote the prices of each block.  We will show that if in GFP
each bidder $i$ truthfully reports her budget and bids $b_i = \min
\{m_i, p_j + \epsilon'\}$, where $j$ is the price block of $i$ in the
PS mechanism, we meet the conditions of the theorem.

Suppose the first price block is determined when bidder $k$ is
considered, and ends at slot $\ell^* \leq k$.  The price $p_1$
satisfies $m_{k+1} \leq p_1 \leq m_k$.  Let $P \subseteq [k]$ denote
the bidders in the first block (the ones in $[k]$ with the $\ell^*$
highest budgets).  Also, we have that all $i \in P$ spend their entire
budget in the PS mechanism, except possibly $k$, who may spend less
than her budget if $m_k = p_1$.  We now argue that GFP will produce
the same allocation as the PS mechanism for this price
block.  For all $i \in P$ we have $b_i = \min \{m_i, p_1 + \epsilon'\}
\geq \min \{m_k, p_1 \} = p_1$.  All bidders $i \in ([k] \setminus P)$
have $b_i \leq p_2 + \epsilon' < p_1$.  All bidders $i \notin [k]$
have $m_i \bidlt m_k$ and so since $b_i \leq m_i$ we get $b_i \bidlt
b_{i'}$ for all $i' \in P$.  We conclude that the bidders in $P$ are
the first to be considered by the GFP mechanism.  Furthermore, if
$k \in P$, and $B_k$ is reduced in the PS mechanism (because $k$ is a threshold bidder), then we must have
$b_k = m_k = p_1$, and so $b_k \bidlt b_i$ for all $i \in P, i \neq
k$.  Thus in this case bidder $k$ is the last bidder in $P$ to be
considered by GFP.  From here it is straightforward to show that GFP
will assign the first $\ell^*$ slots to the bidders in $P$ (almost)
exactly like the PS mechanism does, with at least $c_i -
\epsilon'$ clicks to each $i \in P$; the mechanism will have
$\epsilon'$ clicks left over, which will be assigned to bidders not in
$P$.  Applying this same argument to subsequent price blocks, we
conclude that GFP will assign $c'_i = c_i \pm \epsilon'$ clicks to all
bidders $i$.

To show this is an equilibrium, consider a bidder Alice (call her
``bidder $a$'') that was assigned to price block $j^*$ and received
$c'_a = c_a \pm \epsilon'$ clicks. If Alice spent within $\epsilon'$ of her entire budget,
it means she would not want to raise her bid, since she could not
possibly receive more than $\epsilon'$ additional clicks.  If she did
not spend her budget, then from the observations above we know that
she is bidding her true max-cpc $m_a$, and therefore also does not
want to raise her bid.

It remains to show that Alice does not want to lower her bid.  Let
$\ell_j$ denote the last slot in price block $j$.  Let $P_j$ denote
the set of bidders in price block $j$.  Alice's current bid $b_a$ is
at least $p_j$, and if she keeps her bid above $p_j$ her clicks will
remain $c_a \pm \epsilon$ from the arguments above. 
Let $S = \cup_{j \leq j^*} P_j$.
 If Alice lowers her
bid to $b'_a < p_j$, then all bidders $i \in S$
besides Alice will have $b_i \bidgt b'_a$.  Thus when Alice is considered
by the greedy algorithm, her clicks will be constrained by the
commitments to these bidders.  Furthermore each of these bidders will
still receive at least $c'_i$ clicks.
For all price blocks $j$, we have
$\sum_{i \in P_j} c'_i \geq \sum_{i = \ell_{j-1} + 1}^{\ell_j} D_i - \epsilon'$. 
Thus
$\sum_{i \in S, i \neq a} c'_i \geq (\sum_{i = 1}^{\ell_{j^*}} D_i) - \epsilon' - c'_a$. 
Since $S$ has size $\ell_{j^*}$, this implies that the constraint~\eqref{eq:condition3}
restricts Alice's clicks to at most $c'_a + \epsilon'$.
\end{proofsketchof}

\end{document}